# Photonic Jet with Tunable Focus Based on Water Droplets Freezing from the Outside In

Oleg V. Minin[a], Song Zhou[b,*], and Igor V. Minin[a]

[a]National Research Tomsk Polytechnical University, Tomsk, 634050 Russia

[b]Jiangsu Key Laboratory of Advanced Manufacturing Technology, Faculty of Mechanical and Material Engineering, Huaiyin Institute of Technology, Huai'an, 223003 China

*Corresponding author: zs41080218@126.com

**Abstract**. Water droplets are a perspective highly abundant phase-change material to realize tunable optical lenses. We demonstrated for the first time that freezing mesoscale water droplet could be use as tunable optical lens, such that freezing becomes an asset despite both the low absolute values of the refractive indices of the shell and core materials and their optical contrast. It was shown that the dielectric shell of mesoscale water droplet in the form of solid ice allows controlling both the maximum field intensity and the focus position of the formed photonic nanojet. The formation of ice with air bubbles during the freezing of a water droplet is appropriate for a dynamic increase in the range of change of the focal position compared to solid ice. The proposed concept of a tunnelable spherical lens based on a freezing water drop can be used for microscopy, optical traping in "green" mesotronics.

**Keywords:** water droplet, freezing, photonic jet, tunable focus, mesotronics.

## Introduction

Natural phase-change materials (PCMs) are key photonic materials that have the advantages of a reversible and rapid phase change, scalability and recycling. A water droplet is well known for its ability to focusing a light about 4,000 years ago in the Chow-Foo dynasty in China and from ancient times and is one of those PCM.

Today water droplet is under intensive investigation [1], motivated by bio-comparable green applications [2,3]. Optical resonant effects in water droplet under the light illumination allow understanding the physics of atom [4], design of microlaser [5], sensors [6], etc.

Water-based optical lenses are also still attractive to researchers. In contrast to a solid lens, the fluidic lens allow changing the curvature of the lens, and thus a tunable focal length can be achieved [7-12]. Images based on a droplet lens in front of the camera lens of a cell phone were also demonstrated [13,14]. An attractive feature of water-based optical element is that such lens can be manufactured on demand and recycled after use [15]. Water droplet lens for optical microscopy were studied in [16]. It was concluded that a water droplet lens with millimeter scale diameter and focus length about tens centimeters has a magnification of about 50. An optical microscopes based on moving water droplets as optical lenses were considered in [17,18]. Hemispherical droplet-assisted microscopic imaging technology based on optical tweezer were studied in [19,20]. It has been shown that liquid hemispherical droplets with diameter about of 5 micron can focus light in application to optical microscopy similar to traditional solid spheres.

Silicone coated water droplet based microscopy lens with variable focus was considered in [21,22]. A single water droplet attached to the end of a microliter syringe. The droplet diameter was about millimeter and focal length was about 2-3 of droplet diameter. The shape of the droplet were tuned by applying an external pressure and is associated with its magnification [23]. It has been shown that such water droplet lens coated with a layer of silicone oil is capable of micron-scale resolution imaging.

However, the methods of changing the focal length of a water droplet based lens by changing its shape (volume) are not always acceptable. Besides, water micro-drops has a small volume/area ratio and thus vaporize quickly [24]. In this paper, we propose the concept of a mesoscale lens with a variable focal length based on a freezing water droplet. During the freezing process, water turns into ice, the optical properties of which differ from water. We consider two scenarios: the formation of monolithic ice and ice with air (environment) bubbles.

**Model**

The freezing of spherical droplets with high Bond number in a flow of cold air was considered in [25]. It has been shown that the freezing of drop consists of two main process: cooling of droplet layer down to the phase change temperatures and then the freezing of this spherical layer to ice. At the same time, the shape of a water droplet at a small Bond number can be considered spherical [26], and the water droplets with radii below 50 μm are not to explode due to the expansion of water upon freezing [27].

The dynamics of two-layered water-ice droplet formation by cooled water spherical particles and generation of the photonic nanojet were studied by using the commercial software COMSOL Multiphysics based on the finite elements method. To simplify the simulations without compromising the generality of the results, a cylindrical symmetric model was employed. Considering the cylindrical symmetry, the incident plane wave that propagates along the z-axis with a circular polarization was adopted. The perfect matched layer was applied as the boundary condition. As in [26] the refraction indices of water and solid ice are 1.334 and 1.301, respectively, at the wavelength of $\lambda = 589$ nm. Inside and outside the water–ice drop, the mesh size is less than $\lambda/15$ and $\lambda/8$, respectively. Droplet with initial outer radius of $R=3$ μm, which corresponds to the mesoscale [26, 28] Mie size parameter of $q=\pi D/\lambda \sim 32$ (where $D$ is the diameter of droplet, and $\lambda$ is the radiation wavelength), was immersed in air (n = 1). A schematic diagram is shown in Fig. 1.

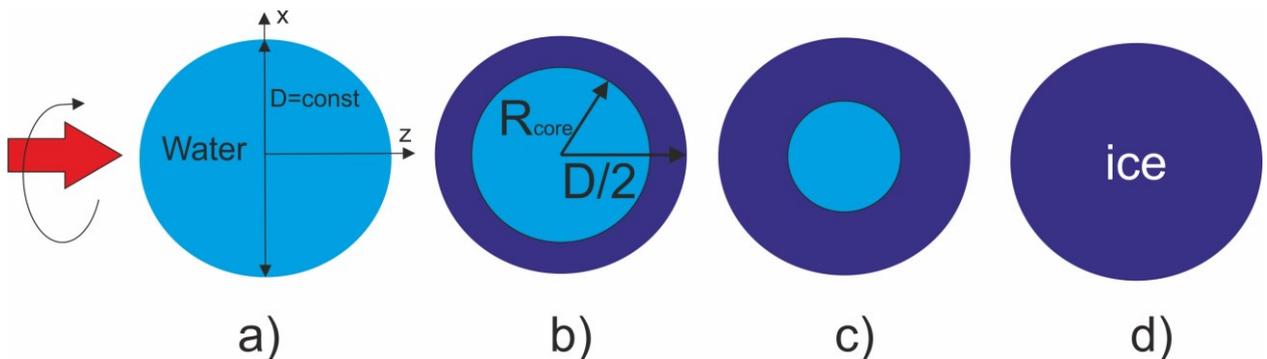

**Figure 1**. Schematic diagram of dynamics of water droplet freezing from the outside in: a) initial water droplet, b)-d) configuration of the water-ice droplet during freezing.

As it was mentioned, in the process of freezing a water droplet, both pure ice and ice with bubbles can form. The gas bubbles dynamics formation accompanying the phase transitions of water were studied in [29-31]. The concept of two-phase water allow described the existence of low- and high-density ice and water [32]. The air bubbles trapped in ice, which result from the dissolved air in liquid water, makes the ice droplet opaque. It was shown [33] that the density of ice with air bubbles is 0.902 of the density of solid ice, assuming that the frozen water drop uniformly captures all the soluted air. Accordingly, the refractive index of ice with bubbles also changes by this coefficient [34,35]. So in the second scenario, we used the effective refractive index $n=1.19$ for ice with air bubbles.

**Simulation results**

To quantitatively characterize the focusing parameters of freezing droplet, we study the following key parameters: the maximal field intensity along the z-direction, the focus position from the droplet center to the point of the maximal field intensity, and the transversal beam waist (FWHM) of the photonic jet in the x-direction at the point of the maximal field intensity. For an optical microscope application, the dynamically variable focus position parameter is of greatest interest to us.

The distribution of the normalized intensity of the electric field $|E/E_0|^2$ in the area of photonic jet formation for various thicknesses of the frozen ice layer is shown in Figure 2. The parameter δ is defined as $R_{core} = (1-\delta)D/2$ to show the thickness of the ice layer.

In a freezing droplet, an illuminating wave first penetrates into the ice (shell) of water-ice droplet. An additional phase shift appears between the wave center (with respect to the main diameter of the particle) and its periphery due to the water (core) is more optically dense than the ice (shell). Due to different refractive index of water and ice, the phase difference formed between different parts of the optical wave leads to the changes the wave front curvature. Since this curvature is positive such deformation of the wavefront favoring the light focusing. The double focusing and defocusing of the incident wave by the shell and core is clear visible in Figure 2a.

One can see from Figures 2-3, that the peak intensity highly increased, while the beam waist of the photonic jet (PNJ) changes only slightly in freezing droplet with ice shell without bubbles. It can be also seen that as the ice shell thickens, the parameters of the photonic jet change non-linearly. This could be due to the destructive and constructive interference of the scattering light of shell and core.

In our case, the shell is made of a less refractive index material (ice) than the core (water), and then the increase in shell thickness affects slightly the PNJ focus position and width. However, in some values of shell thickness the maximal field intensity enhancement is significant. The maximum value of the intensity along the photonic jet and the minimum size of the photonic jet waist are observed near the value of the parameter δ~0.4.

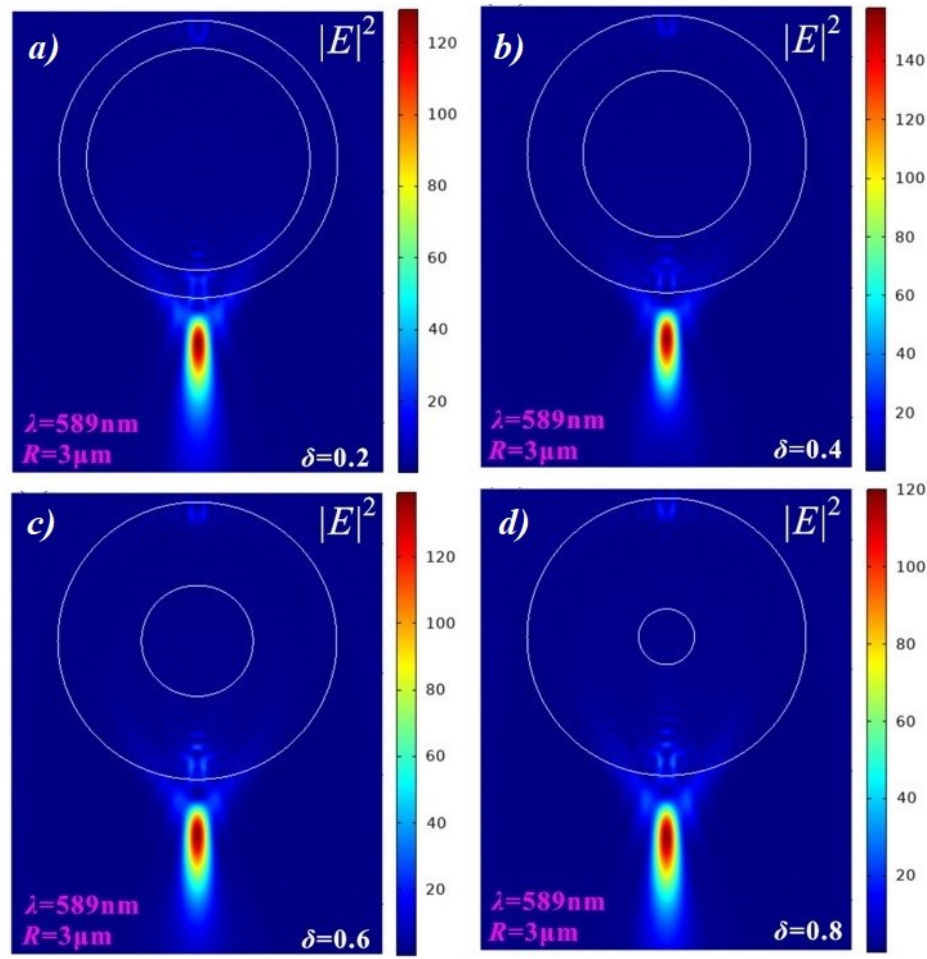

**Figure 2**. Field intensity distributions and PNJ configurations vs parameter δ for ice with $n_{ice} = 1.301$.

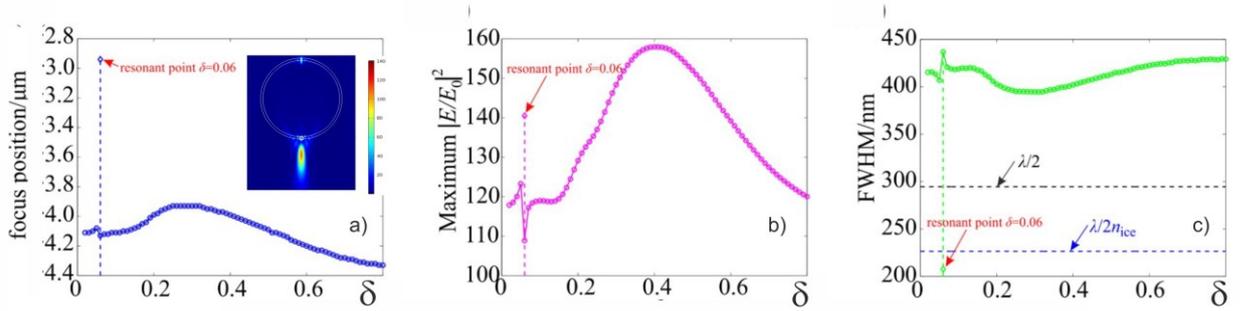

**Figure 3**. Key characteristics of the photonic nanojet generated by freezing droplet: a) focus position, b) maximal field intensity enhancement and c) FWHM vs parameter δ for ice with $n_{ice} = 1.301$.

In the considered case, the value of focus position is found to increase upon increasing the parameter δ till δ~0.3 and the decrease upon further increasing the δ value till 0.8 (Figure 3a). However, the absolute change in the position of the focus is not large and amounts to about 0.4 µm. The opposite situation is observed for beam waist (Figure 3c). The value of FWHM is found to decrease upon increasing the parameter δ till δ~0.3 and the increase upon further increasing the δ value till 0.8. At the same time, the maximum relative field intensity is observed near the value of δ~ 0.4 (Fig.2b), where the maximal electrical field intensity enhancement is about $|E/E_0|^2$~148 (Fig.2b).

Note that for the indicated parameters of the freezing drop, due to the constructive interference of the scattering optical wave of shell and core, Mie and Fano resonances are observed near the value of δ~0.06 or for the thickness of the ice shell of 180 um, where FWHM of the hot spot inside the shell is FWHM=0.36λ. This configuration of the droplet is shown in the insert in Figure 3a. However, a detailed study of resonance effects in this case is beyond the scope of this article and will be considered in a separate paper.

It could be noted that PNJs generated from core–shell dielectric microspheres illuminated by a plane wave, were previously studied, for example, in [36-38]. However, the change in the parameters of the formed PNJ was associated with an increase in the diameter of the sphere due to the thickness of the shell. Moreover, in our case, the refractive index of the core (water) is noticeably lower than was usually considered earlier (about 1.5 and higher), refractive index contrast core-shell is near the unity, and the outer diameter of the sphere (droplet) is constant.

Now let's briefly consider the second scenario - the characteristics of the formed photonic jet, in which ice with an effective refractive index of $n_{ice} = 1.19$, containing air bubbles. Simulation results are shown in Figure 4 below.

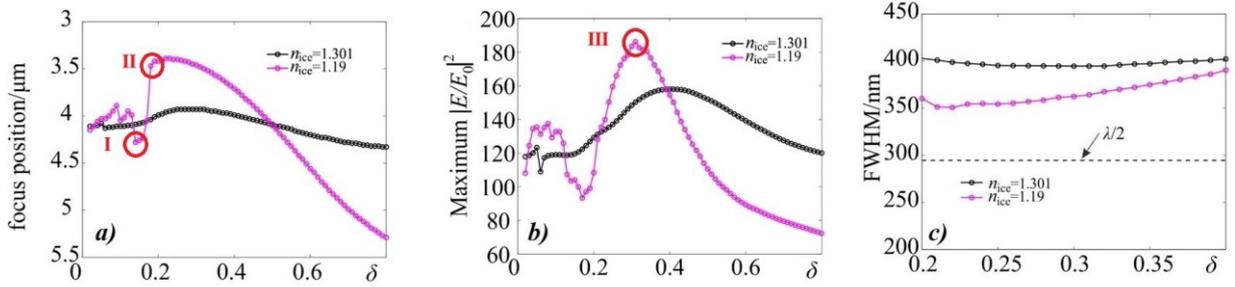

**Figure 4**. Key characteristics of the photonic nanojet generated by freezing droplet: a) focus position, b) maximal field intensity enhancement and c) FWHM vs parameter δ for ice with $n_{ice}=1.19$ (pink) and $n_{ice}=1.301$ (black).

One can see that the dependence of the focal length on the parameter δ becomes more pronounced when the maximum is shifted to the left (Fig.4a). In this case, the range of change in the focal position significantly increases in comparison with solid ice and amounts to about 2 μm, which makes it possible to use the considered effects to control the position of the focus along the optical axis during droplet freezing. But the resonance in the shell (ice) does not arise due to destructive interference of the scattering light of shell and core. At the same time, in the region of the parameter δ~0.2, there is a sharp transition from a local minimum to a maximum in the dependence under consideration (marked with red circles I and II in Fig. 4a).

The dependence of the maximum relative intensity of the electric field $|E/E_0|^2$ on the parameter δ also shifts to the left and becomes more "gradient". Maximum corresponds to the value of δ~0.31 (marked with red circle III in Fig.4b).

The dependence of the beam width of the PNJ on δ for ice with air bubbles ($n_{ice}=1.19$) "shifts" downward (has a smaller value) compared to solid ice ($n_{ice}=1.301$) and varies over a wide ranges which shown in Fig.4c. These features are due to the large optical contrast between the core (water) and shell (ice) materials.

Figure 5 shows the electric field intensity configurations for cases I,II,III, marked in Figure 4 with red circles.

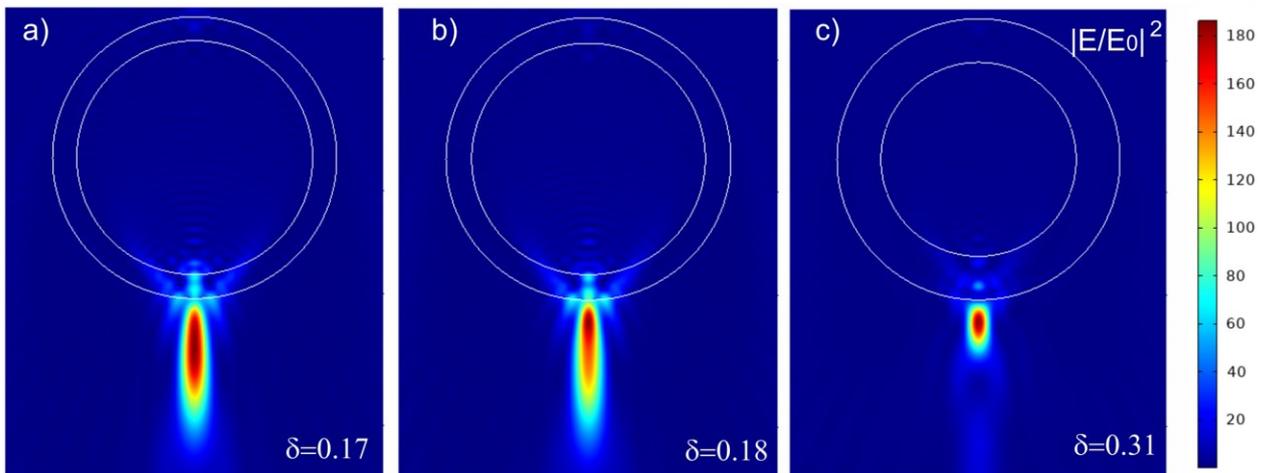

**Figure 5**. Field intensity distributions and PNJ configurations for the cases I,II,III shown in Figure 4, respectively.

One can clear see that at the value of $\delta=0.31$ and $n_{ice}=1.19$ due to lager refractive index contrast, the double focusing of the incident wave by the shell and core is observed. The curvature of the wave front of the light passing through the core increases. Thus, the wave additionally focuses. As a result, the intensity at the focus for $\delta = 0.31$ increases (Fig.4b). At the same time, the length of the PNJ shorten (see Fig. 5c). For δ=0.31 and $n_{ice} = 1.19$, the minimal beam waist of the PNJ is FWHM~ 364 nm (0.618λ).

The double focusing and defocusing of the incident wave by the core and the shell at $\delta = 0.17$ and $\delta = 0.18$ is shown in Figures 5a and 5b, respectively. In the case of $n_{ice} = 1.19$, the range of change in the position of the focus increases significantly to values of about 2 μm or ~3.4 λ in comparison with solid ice with $n_{ice} = 1.301$.

**Conclusion**

The optical properties of water droplets during their freezing process are of interest in a variety of new applications, including temperature control, biomedical applications, temporal photonics [39], etc. The understanding of optical effects at water mesoscale droplet freezing is fundamental importance, which opens new applications of tunable light localization.

We demonstrate the concept of tunable photonic nanojet (temporal PNJ) that is based on a freezing from outside to inside mesoscale water droplet. The feasibility of freezing of a water droplet from outside to inside has been experimentally demonstrated in [27]. It was shown that such freezing droplet makes it possible to focus the light at the shadow side of the droplet into the photonic nanojet with tunable focus position despite both extremely low optical contrast between ice and water and low refractive index of ice and water.

By selecting different scenarios of freezing of a water droplet, including the air bubbles in ice, it is possible to control the key characteristics of the PNJ during freezing. We hope that our work will allow the use of freezing from outside to inside water droplets optical sensors, in optomechanics, and nanoparticle manipulation and trapping [40] in different devices made from strictly natural liquid – "green" mesotronics.